\documentclass[11pt]{article}
\usepackage{graphicx,psfrag,epsf}
\usepackage{bbm}
\usepackage{hyperref}
\usepackage[authoryear]{natbib}
\usepackage[linesnumbered,ruled]{algorithm2e}
\usepackage{algpseudocode}
\usepackage{appendix}
\usepackage[utf8]{inputenc}
\usepackage{amsmath}
\usepackage{amsfonts}
\usepackage{amssymb}
\usepackage{mathtools}  
\usepackage[version=4]{mhchem}
\usepackage{stmaryrd}
\usepackage{bbold}
\usepackage{enumitem}
\usepackage{float}
\usepackage{longtable}
\usepackage{booktabs}
\usepackage{lscape}
\usepackage{setspace}
\setlist[itemize]{label=\textbullet}
\usepackage{subcaption}
\usepackage{authblk}    
\usepackage{hyperref}
\usepackage[margin=0.8in]{geometry}
\usepackage{multirow}
\begin{document}

  \title{\bf Constructing Bayesian optimal designs for discrete choice experiments by simulated annealing}
  \author[1,*]{Yicheng Mao}
  \author[1,2]{Roselinde Kessels}
  \author[1]{Tom van der Zanden}
  \affil[1]{Department of Data Analytics and Digitalization, Maastricht University, P.O. Box 616, 6200 MD Maastricht, The Netherlands}
  \affil[2]{Department of Economics, City Campus, University of Antwerp, Prinsstraat 13, 2000 Antwerp, Belgium}
  \affil[*]{Correspondence: yicheng.mao@maastrichtuniversity.nl }
  \date{} 
  \maketitle
\begin{abstract}
Discrete choice experiments (DCEs) investigate the attributes that influence individuals' choices when selecting among various options. To enhance the quality of the estimated choice models, researchers opt for Bayesian optimal designs that utilize existing information about the attributes' preferences.
Given the nonlinear nature of choice models, the construction of an appropriate design requires efficient algorithms. Among these, the coordinate-exchange (CE) algorithm is commonly employed for constructing designs based on the MNL model. However, as a hill-climbing method, the CE algorithm tends to quickly converge to local optima, potentially limiting the quality of the resulting designs. We propose the use of a simulated annealing (SA) algorithm to construct Bayesian optimal designs. This algorithm accepts both superior and inferior solutions, avoiding premature convergence and allowing a more thorough exploration of potential solutions. Consequently, it ultimately obtains higher-quality choice designs compared to the CE algorithm.
Our work represents the first application of an SA algorithm in constructing Bayesian optimal designs for DCEs. Through extensive computational experiments, we demonstrate that the SA designs generally outperform the CE designs in terms of statistical efficiency, especially when the prior preference information is highly uncertain.

\end{abstract}

\noindent%
\vfill

\newpage

\section{Introduction}\label{Sec:Introduction}
Discrete choice experiments (DCEs) are frequently used to study consumer preferences for the attributes of various goods and have been widely used in fields such as marketing \citep{Rossi2003,Train2009,Liu2024}, health care \citep{BRIDGES2011403,Luyten2015,DEBEKKERGROB20191050}, and transportation \citep{BLIEMER201163,VANACKER2020759}.
Typically, a DCE presents respondents with a group of choice sets containing different alternatives or profiles that are defined by combinations of attribute levels associated with the product or service being studied. By observing and analyzing respondents' selections within these choice sets, researchers can estimate the attractiveness of each attribute and level and further predict consumer behaviour in real-world scenarios.

When planning DCEs, researchers often face the challenge of conducting experiments that can be costly, cumbersome, and time-consuming. 
To obtain accurate parameter estimates and precise predictions with a limited number of observations, researchers frequently employ $\mathcal{D}$-optimal designs \citep{huber1996importance}, which maximize the information matrix of the model under study.
In the context of DCEs, many commonly used software tools, such as Ngene, JMP and the R package idefix, compute the information matrix based on the multinomial logit (MNL) model. This is because the functional form of the MNL model appears in many other more advanced discrete choice models, such as the panel mixed logit model. As a result, $\mathcal{D}$-optimal designs constructed under the MNL framework not only perform well in estimating MNL models, but also in estimating these advanced discrete choice models \citep{BLIEMER2010720}. Additionally, compared to optimizing designs under more complex models, constructing $\mathcal{D}$-optimal designs using the MNL framework requires significantly less computational effort, making it a more efficient and practical choice. 

Given that MNL models are nonlinear in their parameters, computing their information matrix requires the values of these parameters, which are typically unknown at the experimental design stage. To address this issue, \cite{sandor2001designing} proposed the Bayesian optimal design approach, which incorporates prior distributions of the parameters to account for their uncertainty at the experimental design stage.
Through a detailed case study, \cite{Kessels2011usefulness} showed that even for relatively simple MNL models, Bayesian optimal designs generally yield more accurate parameter estimates than orthogonal factorial (utility-neutral) designs across various scenarios.

Since Bayesian optimal designs are difficult to construct theoretically, search algorithms are frequently employed to derive these designs. 
In the context of DCEs, various algorithms have been proposed, such as the modified Fedorov algorithm \citep{cook1980comparison,kessels2006comparison},
the Relabeling and Swapping (RS) algorithm \citep{huber1996importance}, the Relabeling, Swapping, and Cycling (RSC) algorithm \citep{sandor2001designing}, and the coordinate-exchange (CE) algorithm \citep{meyer1995coordinate,kessels2009efficient}. 
Most of the recent work in DCEs has employed the CE algorithm due to its superior performance relative to other methods \citep{Tian2017Efficiency}.
For example, \cite{kessels2009efficient} demonstrated that the CE algorithm not only runs faster than the modified Fedorov algorithm, but also improves the statistical efficiency of the resulting designs.


The CE algorithm begins with the generation of a random starting design and improves this design by evaluating changes on an attribute-by-attribute basis. For each attribute in each profile of the design, the value of the optimality criterion is calculated across all levels of that attribute. A level is only updated if the resulting new design yields a superior criterion value. This procedure is repeated until all the profiles of the design are completed. If an attribute level changes in the current cycle, another complete cycle or iteration through the design is undertaken. This process continues until no changes occur in a complete cycle or until a predefined maximum number of iterations is reached. The candidate-set-free nature of the CE algorithm gives it a significant advantage, particularly in scenarios where profiles contain a large number of attributes or attribute levels \citep{kessels2009efficient}. 

Despite the computational efficiency of the CE algorithm, its hill-climbing approach limits its exploration capacity, making it prone to getting stuck in suboptimal regions of the design space \citep{meyer1995coordinate}. This limitation is particularly problematic when constructing Bayesian optimal experimental designs for DCEs, where the objective function is highly complex and characterized by numerous local optima. As a result, the algorithm is susceptible to premature convergence, which can significantly compromise the quality of the resulting design, thereby reducing the reliability of parameter estimation and prediction.
To mitigate the issue of premature convergence, \cite{kessels2009efficient} proposed initializing the CE algorithm from multiple starting designs. However, this strategy substantially increases the computational cost, making it an inefficient solution.

The aim of this study is to propose a global-search-enhanced simulated annealing (SA) algorithm to address the limitations of the CE algorithm. The SA method was first proposed by \cite{kirkpatrick1983optimization} for finding the global minimum of a cost function that may possess several local minima. For a given objective function, the SA algorithm initiates from a random or specific starting point, coupled with a relatively high initial temperature. As the algorithm progresses, the temperature is gradually decreased until a predetermined termination criterion is reached. At each temperature level, the algorithm generates a new solution by adding a random perturbation to the old solution and accepts this new solution based on specific rules. The probability of acceptance is calculated with reference to the annealing process in metallurgy, which is also the origin of the name ``simulated annealing''. More specifically, if the new solution is superior to the current one, it is accepted. If not, the decision to accept the solution is made based on certain criteria, such as the Metropolis acceptance criterion \citep{metropolis1953equation}. This approach allows for the acceptance of new solutions that may initially worsen the objective function value, particularly at higher temperatures. This flexibility enables the algorithm to escape local optima. As the temperature is slowly lowered, the algorithm is more likely to converge to a global optimum.

The SA algorithm is widely employed in various optimization problems due to its efficiency in globally exploring for more possible solutions, such as in traveling salesman problems \citep{aarts1988quantitative,malek1989serial} and knapsack problems \citep{liu2006improved,qian2007simulated}.
However, its application in the field of optimal experimental design is relatively scarce. To the best of our knowledge, only a few studies, namely \cite{bohachevsky1986generalized}, \cite{meyer1988constructing} and \cite{angelis2001optimal} have delved into this area. Nonetheless, these works primarily focus on optimal exact experimental designs based on Gaussian models. Their approaches have not been applied to choice models, nor to Bayesian optimal designs for these models. 

Our work is the inaugural application of the SA algorithm in constructing Bayesian optimal designs for DCEs. Through extensive computational experiments, we evaluate the performance of both CE and SA algorithms, observing that SA designs generally surpass CE designs, particularly when the prior preference information is highly uncertain.

The rest of the paper is organized as follows. Section \ref{sec:methodology} introduces the MNL model and the Bayesian $\mathcal{D}$-optimal design approach. In Section \ref{sec:SA}, we describe how to apply the SA algorithm to construct Bayesian $\mathcal{D}$-optimal designs, and in Section \ref{sec:parameter}, we provide recommendations for selecting parameters in the SA algorithm.
In Section \ref{Sec:experiments}, we set up computational experiments to evaluate the performance of the SA algorithm against the CE algorithm. 
Finally, Section \ref{Sec:conclusion} discusses the results and future research directions.

\section{The multinomial logit model and Bayesian $\mathcal{D}$-optimality} \label{sec:methodology}
We introduce the MNL model with notations from \cite{Train2009}, and explain how to construct Bayesian $\mathcal{D}$-optimal designs for this model.

\subsection{Multinomial logit model}
The MNL model assumes that respondents to a DCE belong to a target group of decision makers with homogeneous preferences. The model employs random utility theory which describes the utility that a respondent attaches to profile $j$ ($j=1,\dots, J$) in choice set $s$ ($s=1,\dots, S$) as the sum of a systematic and a stochastic component:
\begin{equation}
U_{js} = {\boldsymbol x}^{T}_{js}\boldsymbol{\beta} + \varepsilon_{js}.
\end{equation}
In the systematic component ${\boldsymbol x}^{T}_{js}\boldsymbol{\beta}$, ${\boldsymbol x}_{js}$ is a $m \times 1$ vector containing the attribute levels of profile $j$ in choice set $s$. The vector $\boldsymbol{\beta}$ is a $m \times 1$ vector of parameter values representing the effects of the attribute levels on the utility. This parameter vector is the same for every respondent. The stochastic component $\varepsilon_{js}$ is the error term, which is assumed independently and identically extreme value distributed. Therefore, the MNL probability that a respondent chooses profile $j$ in choice set $s$ is the closed-form expression
\begin{equation}
p_{js} = \frac{\mbox{exp}\left({\boldsymbol x}^{T}_{js}\boldsymbol{\beta}\right)}{\sum_{j=1}^{J}\mbox{exp}\left({\boldsymbol x}^{T}_{js}\boldsymbol{\beta}\right)}, \label{secondmnl}
\end{equation}
where $\boldsymbol{\beta}$ can be estimated using a maximum likelihood approach.

\subsection{Bayesian $\mathcal{D}$-optimal design}
The $\mathcal{D}$-optimality criterion has been most often employed to construct efficient choice designs \citep{huber1996importance,sandor2001designing}. The $\mathcal{D}$-optimality criterion focuses on precise estimation of the parameters by maximizing the determinant of the information matrix related to the model under investigation. Given a design matrix $\bf{X}$ and parameter vector $\boldsymbol{\beta}$, the $\mathcal{D}$-optimality criterion can be defined as
\begin{equation}
\mathcal{D} = \mbox{log} \left|{\bf M}\left({\bf X}, \boldsymbol{\beta}\right)\right|, \label{Dbcriterion}
\end{equation}
where ${\bf M}\left({\bf X}, \boldsymbol{\beta}\right)$ is the information matrix of the parameter estimates. For the MNL model, the information matrix ${\bf M}\left({\bf X}, \boldsymbol{\beta}\right)$ can be obtained as the sum of the information matrices of each of the $S$ choice sets:
\begin{equation}
{\bf M}\left({\bf X}, \boldsymbol{\beta}\right) = \sum_{s=1}^S {\bf X}^{T}_{s}\left({\bf P}_{s} - {\bf p}_{s}{\bf p}^{T}_{s}\right){\bf X}_{s}, \label{info}
\end{equation}
where ${\bf{X}}=({\bf{X}}_{1},\dots,{\bf{X}}_{S})$ represents the model matrix over all choice sets, ${\bf P}_{s} = \mbox{diag}\left({\bf p}_{s}\right)$ with ${\bf p}_{s} = \left(p_{1s},\dots, p_{Js}\right)^{T}$ denoting the MNL probabilities corresponding to all alternatives in choice set $s$. 

Since the information matrix ${\bf M}\left({\bf X}, \boldsymbol{\beta}\right)$ depends on the unknown parameter vector $\boldsymbol{\beta}$ through the choice probabilities, a multivariate prior distribution is often specified for the model parameters, which leads to the Bayesian optimal design \citep{kessels2006comparison}. The Bayesian $\mathcal{D}$-optimality criterion seeks to maximize the  determinant of the information matrix averaged over the prior distribution $\pi(\boldsymbol{\beta})$, and is defined as
\begin{equation}
\mathcal{D}_B = \int_{\mathcal{R}^m} \mbox{log} \left|{\bf M}\left({\bf X}, \boldsymbol{\beta}\right)\right| \pi(\boldsymbol{\beta})\mbox{d}\boldsymbol{\beta}. \label{eq:Dbcriterion}
\end{equation}
Since the formula in Eq. (\ref{eq:Dbcriterion}) has no closed-form solution, the Bayesian $\mathcal{D}$-optimality criterion is usually numerically approximated by taking draws from the prior distribution. In our work, we adopted the sampling methodology developed by \cite{gotwalt2009fast}, which is based on the radial-spherical integration rule initially proposed by \cite{monahan1997spherical}. This method decomposes the integral in Eq. (\ref{eq:Dbcriterion}) into a radial and a spherical surface component. The radial integration leverages the generalized Gauss-Laguerre quadrature, while the spherical integration employs a randomly rotated extended simplex quadrature \citep{mysovskikh1980approximation}. 
\cite{yu2010comparing} demonstrated the superiority of quadrature methods over alternative approaches in assessing the Bayesian optimality criterion.

To compare the performance of two experimental designs with the same number of choice sets and profiles in each choice set, we adopt the relative $\mathcal{D}_B$-efficiency as a measure \citep{Kessels2011usefulness}. 
The Bayesian $\mathcal{D}$-efficiency of a design $\bf X$, relative to another design $\bf X^*$ can be defined as
\begin{equation}\label{eq:D-efficiency}
    \mathcal{D}_{B}\text{-eff}(\mathbf{X},\mathbf{X^*}) = \exp\left(\frac{\mathcal{D}_B(\mathbf{X})-\mathcal{D}_B(\mathbf{X^*})}{\textit{m}}\right),
\end{equation}
where $m$ is the dimension of the parameter vector.

In terms of $\mathcal{D}_B\text{-eff}(\bf{X},\bf{X^*})$, a value of 1 indicates equivalent performance between the two designs. A value greater than 1 suggests that design $\bf X$ outperforms design $\bf X^*$. Conversely, a value smaller than 1 implies superior performance of design $\bf X^*$.


\section{Simulated annealing algorithm}\label{sec:SA}
We describe how we apply the SA algorithm to construct Bayesian $\mathcal{D}$-optimal designs for DCEs. In Section \ref{sec:desSA}, we provide a standard description of the SA algorithm for Bayesian design generation, and in Section \ref{subsec:cooling}, we delve into the details of the SA algorithm's parameter settings.
\subsection{Standard description of the simulated annealing algorithm}\label{sec:desSA}
\begin{algorithm}[h!]
    \caption{Pseudocode for simulated annealing}
    \label{alg:the_alg}
    \SetAlgoLined
    \SetKwInOut{Input}{Input}
    \SetKwInOut{Output}{Output}
    \Input{Initial random choice design $\mathbf{X}$}
    \Output{The best choice design $\mathbf{X}_{Best}$ found by the algorithm}
    Set a value for \textit{Initial Temperature} $T_0$ using a random walk approach\;
    Set iteration counter $k = 0$\;
    Set $\mathbf{X}_{Best} = \mathbf{X}$\;
    Record the objective value $\mathcal{D}_B(\mathbf{X}_{Best})$\;
    \While{Stopping Criterion not met}{
        Update temperature $T_{k}$ according to a \textit{Cooling Function} $f(k, T_0)$\;
        Generate a new design $\mathbf{X'}$ according to an \textit{Exploration Rule}\;
        Compute the Bayesian $\mathcal{D}$-optimality criterion $\mathcal{D}_B(\mathbf{X'})$\;
        Compute acceptance probability $p$:
        \begin{equation}
            p = \min\left\{1, \exp\left(\dfrac{\mathcal{D}_B(\mathbf{X'}) - \mathcal{D}_B(\mathbf{X})}{T_{k}}\right)\right\}\label{eq:prob}
        \end{equation}
        \If{$\mathbf{X'}$ is accepted with probability $p$}{
            Set $\mathbf{X} = \mathbf{X'}$\;
            \If{$\mathcal{D}_B(\mathbf{X}) > \mathcal{D}_B(\mathbf{X}_{Best})$}{
                Update $\mathbf{X}_{Best} = \mathbf{X}$\;

            }
        }
        Increment iteration counter: $k = k + 1$\;
        \If{no new solutions are accepted in the last 1000 iterations}{
            
            Reheat the temperature: $T_{k} = T_0$\;
            Reset iteration counter: $k = 0$\;
        }
    }
    \textbf{return} $\mathbf{X}_{Best}$
\end{algorithm}

The SA algorithm is defined in Algorithm \ref{alg:the_alg}.
This algorithm starts by taking an initial random choice design $\mathbf{X}$ as input. It then proceeds to determine an appropriate value of the Initial Temperature $T_0$ by using a random walk approach. After that, the algorithm sets the number of iterations $k$ to $0$ and memorizes $\mathbf{X}$ as the current best choice design $\mathbf{X}_{Best}$.
Starting from $T_0$, in each iteration, we employ a Cooling Function $f(k,T_0)$ to update the current system temperature $T_{k}$.
At the same time, we create a new choice design $\bf X'$ that is generated from the neighborhood of the current design based on the specified Exploration Rule.
The new design $\bf X'$ is assessed against the current design $\mathbf{X}$ by using the
the Metropolis acceptance criterion as defined in Eq. (\ref{eq:prob}). The main reason we use this criterion is that it will always lead to two cases:
   \begin{description}
    \item[If $\mathcal{D}_{B}(\mathbf{X'}) \geq \mathcal{D}_{B}(\mathbf{X})$:] Then we know that $e^{\frac{\mathcal{D}_{B}(\mathbf{X'}) - \mathcal{D}_{B}(\mathbf{X})}{T_k}} \geq 1$ and $p = 1$. We accept $\mathbf{X'}$ every time, or probability $p$ is 1.
    \item[If $\mathcal{D}_{B}(\mathbf{X'}) < \mathcal{D}_{B}(\mathbf{X})$:] Then we know that $e^{\frac{\mathcal{D}_{B}(\mathbf{X'}) - \mathcal{D}_{B}(\mathbf{X})}{T_k}} < 1$ and $p < 1$. We accept $\mathbf{X'}$ with probability $p$. Note that the ``worse" the value of $\mathcal{D}_{B}(\mathbf{X'})$ compared to $\mathcal{D}_{B}(\mathbf{X})$, the smaller the value of $p$.
\end{description}
When an accepted solution $\bf X'$ also exhibits better performance than the best design so far, it is then adopted as the best design $\mathbf{X}_{Best}$. 
In each iteration, $k$ is increased by $1$, leading to a corresponding decrease in $T_k$ as the iterations progress. Consequently, the probability of the algorithm accepting a worse solution also decreases over time, eventually approaching near zero. 
To prevent premature convergence, we reset the temperature to Initial Temperature $T_0$, when no more new solutions are accepted in the last 1000 iterations. 
The aforementioned process is repeated until the Stopping Criterion is met. In that case, the algorithm terminates and returns the optimal design $\mathbf{X}_{Best}$.

\subsection{Cooling schedule}\label{subsec:cooling}
The parameters that guide the SA algorithm are collectively referred to as the cooling schedule, and are generally the following \citep{franzin2019revisiting}: 
\begin{enumerate}

\item Initial Temperature $T_0$ (line 1 in Algorithm \ref{alg:the_alg}): The initial temperature $T_0$ sets the starting point for the cooling process in the SA algorithm. 
To facilitate an extensive exploration of the solution space, $T_0$ should be large enough to allow for the acceptance of even the worst solutions in the initial phase of the SA algorithm.
Typically, the value of $T_0$ can be adjusted to achieve a predetermined initial acceptance rate \citep{Johnson1989,Tam1992}. According to the Metropolis acceptance criterion as defined in Eq. (\ref{eq:prob}), we have
\begin{equation}
    T_0 = \frac{|\Delta|_{max}}{|\log p_0|},\label{eq: T0}
\end{equation}
where $p_0$ represents the predefined initial probability of accepting a worse solution, which is typically set close to 1. $|\Delta|_{max}$ denotes the maximum absolute gap in the objective function between two consecutive iterations and is often estimated by a random walk approach \citep{BurkardRendl1984}.

Suppose we perform a random walk in the search space and create a sequence of choice designs $\mathbf{X}_0,\mathbf{X}_1,\dots,\mathbf{X}_L$, where $L$ is the length of the random walk. The resulting Bayesian $\mathcal{D}$-optimality criteria can then be defined as $\mathcal{D}_{B}(\mathbf{X}_0),\mathcal{D}_{B}(\mathbf{X}_1),\dots, \mathcal{D}_{B}(\mathbf{X}_L)$. Let $\Delta_{i,i+1} = \mathcal{D}_{B}(\mathbf{X}_{i+1})-\mathcal{D}_{B}(\mathbf{X}_{i})$ represent the difference in the Bayesian $\mathcal{D}$-optimality criteria between the $i$-th and $i+1$-th step. The maximum absolute gap $|\Delta|_{max}$ can then be estimated as
\begin{equation}
    \widehat{|\Delta|_{max}} = \max_{0 \leq i \leq L} |\Delta_{i,i+1}|.\label{eq: deltaavg}
\end{equation}
By incorporating both $\widehat{|\Delta|_{max}}$ and predefined $p_0$ into Eq. (\ref{eq: T0}), an appropriate value for $T_0$ can be determined.
\item Stopping Criterion (line 5 in Algorithm \ref{alg:the_alg}): The stopping criterion governs the termination of the SA algorithm. 
Common stopping criteria include reaching a fixed maximum amount of time \citep{Tam1992,HussinStutzle2014} or a fixed number of iterations for the algorithm's run \citep{Connolly1990}. In this paper, we employ a more flexible adaptive termination criterion: we terminate the algorithm when no new best solution is found during an entire reheating cycle. In contrast to fixed termination criteria, our method based on observations of the actual execution has been implemented and can be flexibly applied in various experimental design contexts.

\item Cooling Function $f(k,T_0)$ (line 6 in Algorithm \ref{alg:the_alg}): The cooling function governs the decrease in temperature. Typically, a cooling function $f(k,T_0)$ should be monotonically decreasing with respect to the number of iterations $k$, thereby progressively reducing the probability of accepting worse solutions. To date, a variety of cooling functions have been extensively employed, such as the geometric cooling function \citep{kirkpatrick1983optimization}, the logarithmic cooling function \citep{geman1984stochastic,strenski1991analysis}, the hyperbolic cooling function \citep{LundyMees1986,SzuHartley1987,Connolly1990}, the and linear cooling function \citep{Dueck1993}. In this paper, we primarily consider two cooling functions: the geometric cooling function and the hyperbolic cooling function. The geometric cooling function is most commonly used and can be represented as
\begin{equation}
    T_{k} = \alpha^{k} T_0,\label{eq:cooling_exp}
\end{equation}
where $\alpha$ denotes the cooling rate that determines how fast the temperature decreases. To ensure that the algorithm thoroughly explores the solution space, $\alpha$ should be chosen to be less than 1 but close to it. In our study, we have set $\alpha = 0.99$. Alternatively, the hyperbolic cooling function is
\begin{equation}
    T_{k} = \frac{T_0}{k+1}.\label{eq:cooling_inv}
\end{equation}
This function decreases the temperature inversely with the iteration number, resulting in a faster cooling rate in the early stages of the algorithm compared to the geometric cooling function, and a slower rate towards the later stages. We refer to Section \ref{sec:parameter} for a more detailed comparison of these two annealing functions.
\item Exploration Rule (line 7 in Algorithm \ref{alg:the_alg}): The exploration rule determines how the SA algorithm searches for a new solution in the neighborhood of the current design at each iteration. In most implementations of SA, a randomly generated neighbor is produced at each iteration \citep{franzin2019revisiting}. In the context of optimal design for DCEs, we propose that, in each iteration, the newly generated random neighbor differs from the current design in at most one profile within a single choice set. There are two reasons for this approach. First, this exploration rule introduces only a small random perturbation to the current design at each iteration, which is beneficial for the algorithm's repeated search and helps prevent drastic changes that could hinder convergence. Second, this method reduces computational burden. If we change only one choice set in each iteration, we can utilize the following formula to update the information matrix of a newly generated choice design $\mathbf{X^*}$:
\begin{equation}
   {\bf M}\left({\bf X^*}, \boldsymbol{\beta}\right) = {\bf M}\left({\bf X}, \boldsymbol{\beta}\right) -{\bf X}^{T}_{s}\left({\bf P}_{s} - {\bf p}_{s}{\bf p}^{T}_{s}\right){\bf X}_{s} +{\bf X}^{*\,T}_{s}\left({\bf P}^*_{s} - {\bf p}^*_{s}{\bf p}^{*\,T}_{s}\right){\bf X}_{s}^*, 
\end{equation}
where ${\bf M}({\bf X}, \boldsymbol{\beta})$ represents the information matrix for the current design ${\bf X}$, and ${\bf M}({\bf X^*}, \boldsymbol{\beta})$ is the updated information matrix for the newly generated design ${\bf X^*}$. 
 ${\bf P}^*_{s}$ and ${\bf p}^*_{s}$ can be calculated based on the specific choice set ${\bf X}_{s}^*$ added in the current iteration.
 By updating the information matrix incrementally, adjusting only for changes in one choice set rather than recalculating the entire matrix, the computational burden is significantly reduced. This advantage becomes particularly evident when the number of choice sets is large. 
 Based on the approach of modifying one choice set per iteration, we focus on two exploration rules. The first rule is attribute-based: in each iteration, we alter the level of a single attribute within one profile of the current design, introducing minimal perturbation. The second rule is profile-based: in each iteration, we randomly select one profile from a choice set and randomly transform it into another possible profile, allowing for changes in the levels of all attributes within that profile. A detailed comparison of the performance of these two exploration rules can be found in Section \ref{sec:parameter}.
\end{enumerate}
\section{Parameter selection for the cooling schedule} \label{sec:parameter}
The selection of parameters for the cooling schedule has a significant impact on the performance of the SA algorithm. We focus specifically on two critical components of the cooling schedule: the cooling function and the exploration rule. Through an extensive simulation experiment, we compare the performance of different parameter settings and provide recommendations for practical applications.

\subsection{Simulation setup}\label{subsec:cooling_setting}
Our simulation study considers two types of cooling functions: the geometric cooling function, as presented in Eq. (\ref{eq:cooling_exp}), and the hyperbolic cooling function, given in Eq. (\ref{eq:cooling_inv}). Additionally, we examine two different exploration rules, namely the attribute-based and profile-based approaches described in Section \ref{subsec:cooling}. By combining these options, we obtain a total of four distinct cooling schedules for evaluation.

We applied each cooling schedule to a DCE consisting of 15 choice sets, with 2 profiles in each choice set. The experiment considers 6 attributes, where the first three attributes are 2-level attributes, and the last three 3-level attributes. Our approach utilizes effects-type coding for the attributes which ensures that the sum of the levels for each attribute equals zero, necessitating the estimation of the coefficients for all but the last level of each attribute. As a result, the dimension $m$ of the parameter $\boldsymbol{\beta}$ in the MNL model is $2+2+2+3+3+3-6=9$.

Bayesian optimal designs require the specification of a prior distribution for the parameter values, which includes both a mean vector $\boldsymbol{\beta_0}$ and a variance-covariance matrix $\boldsymbol{\Sigma_0}$. 
In this simulation study, we consider a commonly used prior distribution in practical applications, a naive prior distribution, where the prior mean is assumed to be 0 and the prior variance is 1.

To ensure robustness of the results, we generated 100 random initial designs as input to the SA algorithm. For each initial design, we applied the four different cooling schedules and recorded both the runtime at the end of the algorithm and the Bayesian $\mathcal{D}$-optimality criterion value of the final design.

For the initial temperature of the SA algorithm, we utilized the random walk approach introduced in Section \ref{subsec:cooling}. Specifically, in each case, we performed a random walk within the solution space with a length of 100 to estimate $|\Delta|_{max}$. As for the initial probability $p_0$ of accepting worse solutions, we set it at 0.99. By inserting these values into Eq. (\ref{eq: T0}), we determined the corresponding initial temperature $T_0$.

All code was developed using MATLAB R2023b and executed on a system equipped with an 11th Gen Intel(R) Core(TM) i5-1135G7 processor, running at 2.40 GHz (with a turbo boost up to 2.42 GHz).

\subsection{Cooling schedule evaluation}
Table \ref{tab:parameter selection} presents a comparison of the four cooling schedules. The schedules were evaluated based on 100 random initial designs, and results are summarized in terms of the average Bayesian $\mathcal{D}$-optimality criterion values and the average runtimes. It is clear that the average $\mathcal{D}_B$-value for the hyperbolic cooling function is higher than that for the geometric cooling function. In particular, when combined with the attribute-based exploration rule, the hyperbolic cooling function achieves the largest average $\mathcal{D}_B$-value of 5.93. In contrast, the combination of the geometric cooling function with the profile-based exploration rule achieves a $\mathcal{D}_B$-efficiency of 94.70\% compared to the average $\mathcal{D}_B$-value of the best cooling schedule. 

However, it is important to note that higher $\mathcal{D}_B$-values come at the cost of increased computational time. The runtime for the hyperbolic cooling function is generally longer than that of the geometric cooling function. This difference is inherent to the nature of these functions: the geometric cooling function tends to reach lower temperatures in fewer iterations, leading to a rapid reduction in the probability of accepting worse solutions, and consequently, shorter runtimes. The profile-based exploration rule performs relatively worse, showing lower average $\mathcal{D}_B$-values compared to the attribute-based exploration rule under the same cooling function conditions. Additionally, the profile-based rule tends to require longer runtimes.

\begin{table}[h!]
    \centering
        \caption{Performance comparison of four cooling schedules with 100 random starting points.}
    \begin{tabular}{cccc}
    \hline
           Cooling function&  Exploration rule&  Avg. $\mathcal{D}_B$& Avg. runtime (s)\\
    \hline
         Geometric&  Profile&  5.44& \phantom{1}44.33\\
         Geometric&  Attribute&   5.55& \phantom{1}23.74\\
         Hyperbolic &  Profile&   5.75& 111.74\\
         Hyperbolic &  Attribute&  5.93& \phantom{1}82.81\\
    \hline
    \end{tabular}

    \label{tab:parameter selection}
\end{table} 

\begin{figure}
\centering
\includegraphics[width=\textwidth]{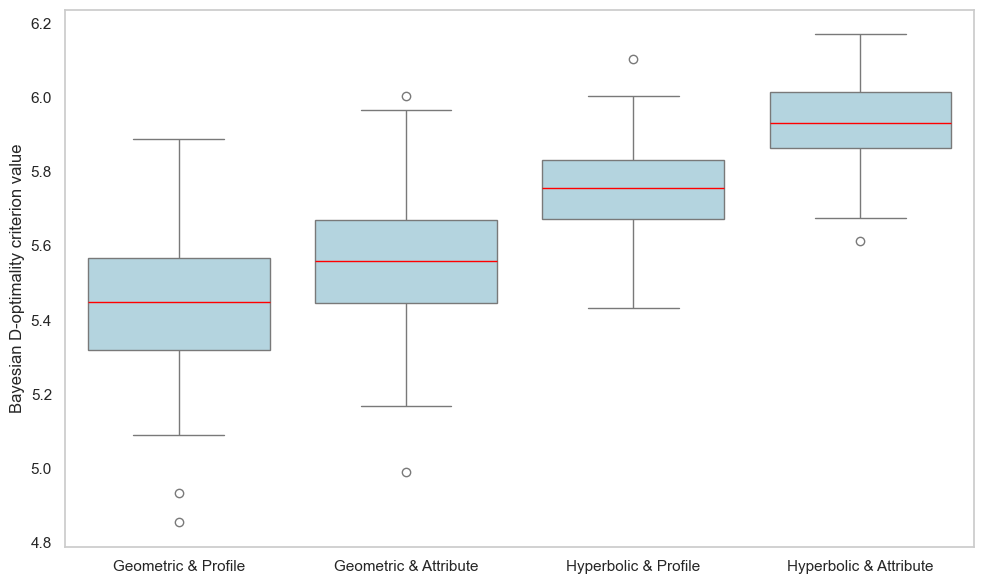}
\caption{Comparison of Bayesian $\mathcal{D}$-optimality criterion values of designs across different cooling schedules.}
\label{fig:cooling_schedule}
\end{figure}

Figure \ref{fig:cooling_schedule} presents a boxplot comparison of the Bayesian $\mathcal{D}$-optimality criterion values of the final designs obtained across the four cooling schedules. The red line within each box indicates the median $\mathcal{D}_B$-value for each cooling schedule. The figure clearly demonstrates that the combination of the hyperbolic cooling function and the attribute-based exploration rule achieves a higher median $\mathcal{D}_B$-value compared to the other schedules. Furthermore, the smaller variance in the $\mathcal{D}_B$-values for this combination suggests greater stability in performance across different random starting points.

We conducted a two-way ANOVA to assess the effects of the cooling function and exploration rule on the final $\mathcal{D}_B$-values. Both factors, the Cooling Function and the Exploration Rule, exhibited highly significant main effects on the $\mathcal{D}_B$ values (both $p < 0.0001$), indicating that each independently influences the quality of the design. Additionally, the interaction between the Cooling Function and Exploration Rule was statistically significant, though to a lesser extent ($p = 0.018$). Further analysis using Tukey's HSD test confirmed that the combination of the hyperbolic cooling function and the attribute-based exploration rule achieved the highest mean $\mathcal{D}_B$-value. Given that the primary objective of this study is to improve design quality, we will utilize this cooling schedule in the subsequent computational experiments.

Based on these results, we recommend that practitioners adopt the attribute-based exploration rule, as it offers advantages over the profile-based exploration rule in both runtime efficiency and design quality. The hyperbolic cooling function is recommended when achieving higher design quality is the priority and longer runtimes are acceptable. Conversely, for cases where computational resources are limited, the geometric cooling function provides a reasonable trade-off between quality and runtime.

\section{Computational experiments}\label{Sec:experiments}
In this section, we evaluate the performance of the CE and SA algorithms through two sets of computational experiments. The first set focuses on assessing the relative $\mathcal{D}_B$-efficiency of the optimal choice designs generated by both algorithms under various experimental settings. In the second set, we simulate the choices of 100 respondents using CE and SA optimal designs to compare the performance of the designs in terms of estimation accuracy, prediction accuracy, and sample size efficiency.

\subsection{Bayesian $\mathcal{D}$-efficiency comparison}\label{subsec:simulation1}
\subsubsection{Experimental setup}
We assess the performance of the CE and SA algorithms in nine distinct experimental setups, each characterized by different prior information settings. The design problem involves constructing a choice design consisting of 15 choice sets, each with 2 profiles.
In each experiment, we deal with the same six attributes as introduced in \ref{subsec:cooling_setting}, employing effects-type coding for the attributes.

Bayesian optimal designs necessitate the specification of a prior mean vector $\boldsymbol{\beta_0}$ and a prior variance-covariance matrix $\boldsymbol{\Sigma_0}$. For $\boldsymbol{\beta_0}$, we assume that the attractiveness of each attribute increases with the level, with the last level being the most preferred and the first level being the least preferred. We define $\boldsymbol{\beta_0}$ as a function of a scale parameter $\lambda$:
\begin{equation}
    \boldsymbol{\beta}_{0}(\lambda)=(-\lambda, -\lambda, -\lambda, -\lambda, 0, -\lambda, 0,-\lambda, 0)^{T},
\end{equation} 
where $\lambda \in \{1,\frac{1}{2},\frac{1}{3}\}$. Similarly, we set $\boldsymbol{\Sigma_0}$ as a function of a scale parameter $\kappa$:

\begin{equation}
    \boldsymbol{\Sigma}_0(\kappa) = \begin{pmatrix}
\kappa^2 & 0 & 0 & 0 & 0 & 0 & 0 & 0 & 0 \\
0 & \kappa^2 & 0 & 0 & 0 & 0 & 0 & 0 & 0 \\
0 & 0 & \kappa^2 & 0 & 0 & 0 & 0 & 0 & 0 \\
0 & 0 & 0 & \kappa^2 & -0.5\kappa^2 & 0 & 0 & 0 & 0 \\
0 & 0 & 0 & -0.5\kappa^2 & \kappa^2 & 0 & 0 & 0 & 0 \\
0 & 0 & 0 & 0 & 0 & \kappa^2 & -0.5\kappa^2 & 0 & 0 \\
0 & 0 & 0 & 0 & 0 & -0.5\kappa^2 & \kappa^2 & 0 & 0 \\
0 & 0 & 0 & 0 & 0 & 0 & 0 & \kappa^2 & -0.5\kappa^2 \\
0 & 0 & 0 & 0 & 0 & 0 & 0 & -0.5\kappa^2 & \kappa^2 \\
\end{pmatrix},
\end{equation}
where $\kappa \in \{1,\frac{1}{2},\frac{1}{3}\}$. The larger the value of $\kappa$, the larger the uncertainty about the prior mean. For the 3-level attributes, we ensure that the parameters of the first two levels are negatively correlated with each other, so that the variances of all parameters, especially the third-level parameter, are all equal \citep{kessels2008recommendations}. 
By varying the values of $\lambda$ and $\kappa$, we establish nine distinct prior distributions, which can cover most scenarios that occur in real-life experiments. 

For a robust comparison, we generated 100 random starting points for each prior distribution, which served as input for both the CE and SA algorithms. For the CE algorithm, the termination criterion was set such that the $\mathcal{D}_B$-value did not improve within an entire cycle, ensuring that the CE algorithm fully converged and yielded the optimal design for each starting point.

For the SA algorithm's cooling schedule, we selected the hyperbolic cooling function and the attribute-based exploration rule based on design quality considerations. The settings for the other parameters of the SA algorithm are consistent with those outlined in Section \ref{subsec:cooling_setting}.

For each random starting point, we recorded the final $\mathcal{D}_B$-value of the choice design produced by both the CE and SA algorithms, along with the corresponding runtime.

\subsubsection{Results}
\begin{figure}[H] 
    \centering
    \begin{subfigure}{0.3\textwidth}
        \includegraphics[width=\linewidth]{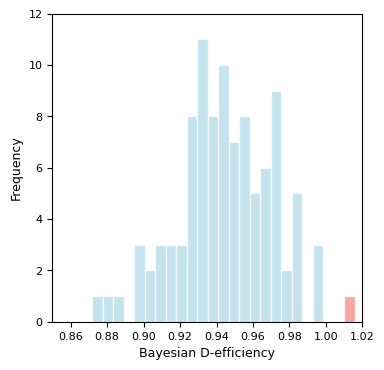} 
        \caption{$\lambda = 1, \kappa = 1$}
    \end{subfigure}
    \begin{subfigure}{0.3\textwidth}
        \includegraphics[width=\linewidth]{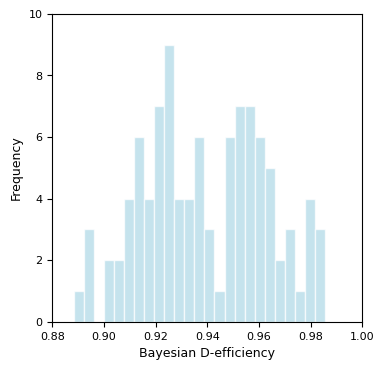}
        \caption{$\lambda = 1, \kappa = 1/2$}
    \end{subfigure}
    \begin{subfigure}{0.3\textwidth}
        \includegraphics[width=\linewidth]{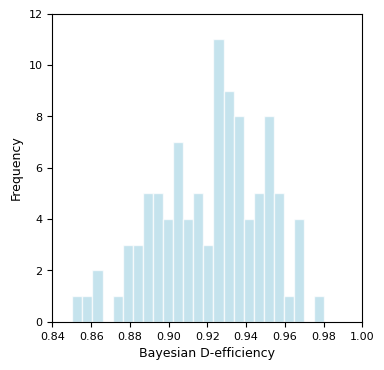}
        \caption{$\lambda = 1, \kappa = 1/3$}
    \end{subfigure}

    \begin{subfigure}{0.3\textwidth}
        \includegraphics[width=\linewidth]{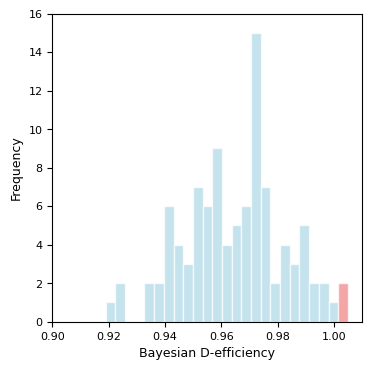}
        \caption{$\lambda = 1/2, \kappa = 1$}
    \end{subfigure}
    \begin{subfigure}{0.3\textwidth}
        \includegraphics[width=\linewidth]{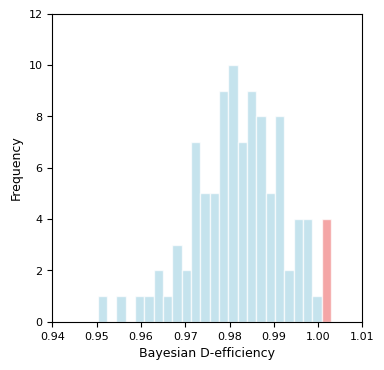}
        \caption{$\lambda =1/2, \kappa =1/2$}
    \end{subfigure}
    \begin{subfigure}{0.3\textwidth}
        \includegraphics[width=\linewidth]{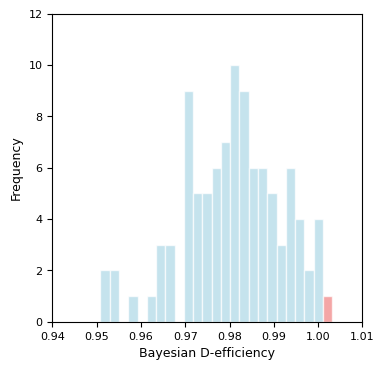}
        \caption{$\lambda = 1/2, \kappa = 1/3$}
    \end{subfigure}

    \begin{subfigure}{0.3\textwidth}
        \includegraphics[width=\linewidth]{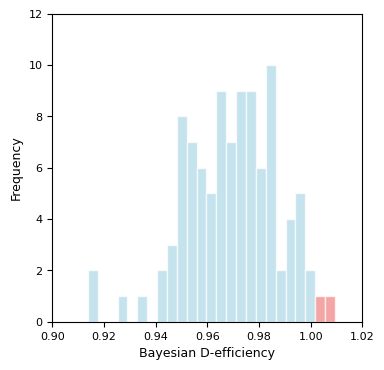}
        \caption{$\lambda = 1/3, \kappa = 1$}
    \end{subfigure}
    \begin{subfigure}{0.3\textwidth}
        \includegraphics[width=\linewidth]{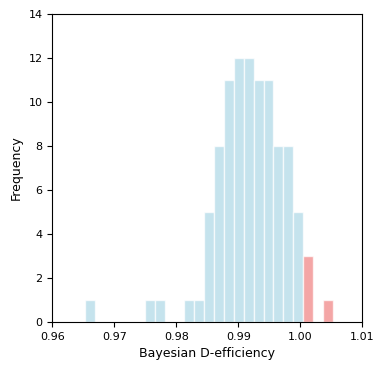}
        \caption{$\lambda = 1/3, \kappa = 1/2$}
    \end{subfigure}
    \begin{subfigure}{0.3\textwidth}
        \includegraphics[width=\linewidth]{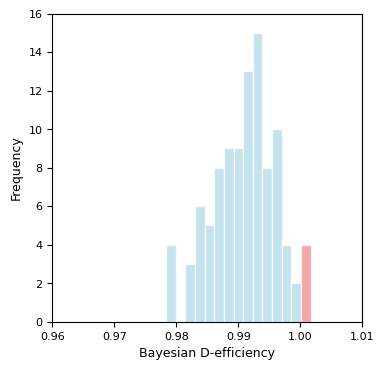}
        \caption{$\lambda = 1/3, \kappa = 1/3$}
    \end{subfigure}
    
   \caption{Histograms of Bayesian $\mathcal{D}_{B}$-efficiencies of CE optimal designs relative to SA optimal designs (or $\mathcal{D}_{B}\text{-eff}({\bf {X}}_{CE},{\bf {X}}_{SA})$) across nine different prior distributions.}
\label{fig:CE_VS_SA_2}
\end{figure}

Figure \ref{fig:CE_VS_SA_2} presents nine histograms showing the distributions of Bayesian $\mathcal{D}_{B}$-efficiencies, $\mathcal{D}_{B}\text{-eff}({\bf {X}}_{CE},{\bf {X}}_{SA})$, of 100 CE optimal designs relative to 100 SA optimal designs across nine different prior distributions. The red bars in each subplot represent instances where the relative Bayesian $\mathcal{D}_{B}$-efficiency exceeds 1 under the corresponding prior distribution. As shown in the figure, it is clear that regardless of the prior distribution, the majority of relative Bayesian $\mathcal{D}$-efficiencies are smaller than 1. This indicates that the SA designs mostly outperform the CE designs.

\begin{table}[h!]
    \centering
     \caption{Average $\mathcal{D}_B$-values, runtimes, and relative $\mathcal{D}_B$-efficiencies for the CE and SA algorithms across different $\lambda$ and $\kappa$ values based on 100 random initial designs.}
    \begin{tabular}{llcccccc}
    \hline
\multirow{2}{*}{$\lambda$} & \multirow{2}{*}{$\kappa$}& \multicolumn{2}{c}{CE}&  &\multicolumn{2}{c}{SA}&\multirow{2}{*}{Avg. $\mathcal{D}_{B}\text{-eff}({\bf {X}}_{CE},{\bf {X}}_{SA})$}\\
\cline{3-4} \cline{6-7}
  & &  Avg. $\mathcal{D}_B$& Avg. runtime (s)& & Avg. $\mathcal{D}_B$&  Avg. runtime (s)& \\
  \hline
           1&  1
&  \phantom{1}2.43 &  8.50 &&  \phantom{1}2.94 &  \phantom{1}86.45 & 94.50\%\\
           1&  0.5&  12.54 &  7.88 &&  13.11 &  \phantom{1}94.58 & 93.91\%\\
           1&  0.33&  13.86 &  7.30 &&  14.59 &  \phantom{1}85.11 & 92.15\%\\
           0.5&  1
&  \phantom{1}3.95 &  8.71 &&  \phantom{1}4.28 &  \phantom{1}89.27 & 96.46\%\\
           0.5&  0.5&  14.49 &  7.31 &&  14.65 &  \phantom{1}99.56 & 98.21\%\\
           0.5&  0.33&  16.03 &  7.46 &&  16.21 &  \phantom{1}89.71 & 98.07\%\\
           0.33&  1
&  \phantom{1}4.57 &  8.45 &&  \phantom{1}4.85 &  \phantom{1}84.73 & 96.92\%\\
           0.33&  0.5&  15.29 &  6.50 &&  15.37 &  107.17 & 99.18\%\\
           0.33&  0.33&  16.88 &  6.33 &&  16.96 &  107.37 & 99.10\%\\
           \hline
    \end{tabular}
   
    \label{tab:CE_VS_SA_2}
\end{table}

Table \ref{tab:CE_VS_SA_2} presents the average Bayesian $\mathcal{D}_B$-values, runtimes, and relative $\mathcal{D}_B$-efficiencies of the 100 optimal designs generated by the CE and SA algorithms under nine different prior distributions.
To further compare the optimal designs, we conducted a Wilcoxon signed-rank test on the average $\mathcal{D}_B$-values. The results indicate that the average $\mathcal{D}_B$-values of the SA designs are significantly higher than those of the CE designs ($p < 0.0001$), demonstrating the superior design-generating quality of SA. This difference is particularly pronounced in scenarios with large prior variances, while in scenarios with smaller prior variances, the performances of the two algorithms are more comparable.

The likely reason for these findings is that when the prior variance is larger, the objective function becomes more complex, with a higher likelihood of containing multiple local optima. In such cases, the hill-climbing nature of the CE algorithm makes it more prone to early convergence to one of these local optima. In contrast, the SA algorithm can escape local optima due to its inherent stochastic nature, accepting occasionally worse solutions. This characteristic increases the likelihood that the SA algorithm does not get entangled in suboptimal local solutions, allowing it to explore a broader range of potential solutions. As a result, the SA algorithm is more capable of identifying better choice designs, particularly when dealing with complex objective functions where the risk of local optima is high. However, as a trade-off for producing higher-quality designs, the runtime of the SA algorithm is longer than that of the CE algorithm, often about 10 times longer in various cases. Nevertheless, its runtime remains within a practically feasible range, making it a viable option in practice.

The specific settings of the experiments can affect the performance of the algorithms. Therefore, we also considered scenarios where each choice set comprises three profiles. The results obtained in these scenarios are similar to those with two profiles. Further details and comparisons can be found in Table \ref{tab:CE_VS_SA_3} and Figure \ref{fig:CE_VS_SA_3} of the Appendix.

\subsection{Estimation accuracy, prediction accuracy, and sample size efficiency comparison}
\subsubsection{Experimental setup}
We assess the performance of the optimal designs generated by the CE and SA algorithms in terms of estimation accuracy, prediction accuracy, and sample size efficiency.

In practice, researchers often run the CE algorithm from multiple random initial points and select the best design as the final design \citep{kessels2009efficient,goos2011optimal}. Therefore, in this experiment, we first run the CE algorithm with 100 random starting points, using a termination criterion where the algorithm stops if no better design is found within a complete cycle. The design with the largest $\mathcal{D}_B$-value among the 100 random starting points is chosen as the final output of the CE algorithm. For a fair comparison, the SA algorithm is allocated the same runtime as the CE algorithm. Specifically, we record the runtime of the CE algorithm and use it as the maximum runtime for the SA algorithm, which serves as its termination criterion.

We applied both algorithms to the optimal design of a DCE with 15 choice sets, each consisting of two profiles. This DCE includes six attributes where the first three attributes are 2-level attributes, and the last three 4-level attributes. The attributes are coded using effects-type coding. Regarding the prior information needed to construct the Bayesian $\mathcal{D}$-optimal designs, we set the prior mean $\boldsymbol{\beta_0}$ to $(-1,-1,-1,-1,-0.5,0.5,-1,-0.5,0.5,-1,-0.5,0.5)^{T}$ and the prior covariance matrix $\boldsymbol{\Sigma_0}$ to the identity matrix ${\bf I}_{12}$.

After obtaining the optimal choice designs from both algorithms, we simulated responses for each optimal design by assuming that $\boldsymbol{\beta_{0}}$ is the true parameter vector $\boldsymbol{\beta^{*}}$. To compare the performance of the designs in terms of estimation and prediction accuracy, we assumed that 100 respondents participated in this DCE. We applied an MNL model to analyze the choice data for each design and recorded the parameter estimates. To ensure robustness of the results, we repeated this simulation 1000 times, resulting in 1000 simulated datasets and corresponding parameter estimates for each design.

Moreover, as a third comparison measure, we evaluated the minimum sample size required for the CE design and the SA design to achieve statistically significant parameter estimates. Therefore, we considered 100 scenarios where the number of participants ranged from 1 to 100. For each scenario, we generated 1000 simulated datasets with the true parameter vector $\boldsymbol{\beta^{*}}$. We then estimated an MNL model for each simulated dataset and recorded the $t$-ratios for the part-worth estimates.

\subsubsection{Results}
The runtime for both the CE and SA algorithms was 1696.6 seconds. The optimal choice designs from the algorithms can be found in Table \ref{tab:design2} of the Appendix. The SA design outperformed the CE design in terms of the $\mathcal{D}_B$-optimality criterion, with 
a relative $\mathcal{D}_{B}\text{-eff}({\bf {X}}_{CE},{\bf {X}}_{SA})$ of 95.87\%.

To evaluate the accuracy of the parameter estimates of the CE and SA designs, we employed the Expected Mean Square Error (EMSE) of the estimated parameters, which can be expressed as

\begin{equation}\label{eq:EMSEbeta}
    \text{EMSE}_{\hat{\boldsymbol{\beta}}}(\boldsymbol{\beta^{*}}) =  
    \int_{\mathcal{R}^{k}} 
    \left(\hat{\boldsymbol{\beta}} - \boldsymbol{\beta^{*}}\right)^{T}
    \left(\hat{\boldsymbol{\beta}} - \boldsymbol{\beta^{*}}\right)
    \pi(\hat{\boldsymbol{\beta}}) \, \mathrm{d}\hat{\boldsymbol{\beta}},
\end{equation}
where $\pi(\hat{\boldsymbol{\beta}})$ is the distribution of the estimates. A smaller $\text{EMSE}_{\hat{\boldsymbol{\beta}}}(\boldsymbol{\beta^{*}})$ value indicates greater accuracy of the estimated parameters. 
Following the approach in \cite{yu2008model}, the EMSE value is approximated by
\begin{equation}\label{eq:EMSEbeta2}
    \text{EMSE}_{\hat{\boldsymbol{\beta}}}(\boldsymbol{\beta^{*}}) =  \frac{1}{N} \sum_{n=1}^{N} 
    \left(\hat{\boldsymbol{\beta}}^{n} - \boldsymbol{\beta^{*}}\right)^{T}
    \left(\hat{\boldsymbol{\beta}}^{n} - \boldsymbol{\beta^{*}}\right),
\end{equation}
where $N=1000$ represents the number of simulations, and $\hat{\boldsymbol{\beta}}^{n}$ denotes the vector of estimates obtained from the $n$-th simulated dataset.

Similarly, we employed the EMSE to evaluate the performance of the designs in terms of prediction accuracy. The term $\text{EMSE}_{\hat{\mathbf{p}}}(\boldsymbol{\beta^{*}})$
relates to the predicted probabilities for the complete design that encompasses all $Q$ possible choice sets of size $J$.
Therefore, we define $\text{EMSE}_{\hat{\mathbf{p}}}(\boldsymbol{\beta^{*}})$ as
\begin{equation}\label{eq:EMSEp1}
\text{EMSE}_{\hat{\mathbf{p}}}(\boldsymbol{\beta}^{*}) =  
\frac{1}{N \cdot J \cdot Q} \sum_{n=1}^{N} 
\left(\hat{\mathbf{p}}(\hat{\boldsymbol{\beta}}^{n}) - \mathbf{p}\left(\boldsymbol{\beta}^{*}\right)\right)^{T}
\left(\hat{\mathbf{p}}(\hat{\boldsymbol{\beta}}^{n}) - \mathbf{p}\left(\boldsymbol{\beta}^{*}\right)\right),
\end{equation}
where $N=1000$ represents the number of simulations, $\bf{p}(\boldsymbol{\beta^{*}})$ is the vector of true MNL probabilities in the complete choice design, and $\hat{\mathbf{p}}(\hat{\boldsymbol{\beta}}^{n})$ is the vector of predicted probabilities using the estimates $\hat{\boldsymbol{\beta}}^{n}$ from the $n$-th simulated dataset. As in \cite{kessels2006comparison}, we considered all possible choice sets of size two, which means $J = 2$. Given that the design problem involves \(2 \times 2 \times 2 \times 4 \times 4 \times 4 = 512\) different alternatives, there are $Q = \binom{512}{2} = 130816$ possible choice sets.
A smaller value of $\text{EMSE}_{\hat{\mathbf{p}}}(\boldsymbol{\beta^{*}})$ indicates more accurate predictions of the probabilities.

\begin{table}[h]
  \caption{EMSE values comparing estimation and prediction accuracy of the CE and SA designs.}
    \centering
    \begin{tabular}{ccc}
    \hline
         &  CE& SA\\
    \hline
         $\text{EMSE}_{\hat{\boldsymbol{\beta}}}$&  0.08835& 0.08362\\
         $\text{EMSE}_{\hat{\bf{p}}}$&  0.00112& 0.00099\\
    \hline
    \end{tabular}
  
    \label{tab:EMSE}
\end{table}

Table \ref{tab:EMSE} displays the EMSE values of the CE and SA designs. Results indicate that, in terms of both estimation accuracy and prediction accuracy, the EMSE values of the SA design are lower. Specifically, for estimation accuracy, the EMSE value of the CE design increased by approximately 6\% compared to the SA design, while for prediction accuracy, the EMSE value of the CE design increased by about 13\%. This suggests that the SA design achieves more precise parameter estimates and predictions compared to the CE design.

In addition to estimation and prediction accuracy, practitioners often focus on the sample size efficiency ($\mathcal{S}$-efficiency) of a choice design \citep{rose2013sample}. $\mathcal{S}$-efficiency evaluates whether a choice design can yield statistically significant parameter estimates with a smaller sample size requirement. Unlike $\mathcal{D}$-efficiency, which assesses the overall precision of parameter estimates, $\mathcal{S}$-efficiency specifically focuses on the most difficult parameter to estimate. To measure $\mathcal{S}$-efficiency, we use the minimum absolute $t$-ratio introduced in \cite{rose2013sample} and defined by
\begin{equation}\label{eq:t-ratio}
    |t|_{\text{min}} = \min_k \left\{ \left| \frac{\hat{\beta}_k}{\text{se}(\hat{\beta}_k)} \right| \right\},
\end{equation}
where $\hat{\beta}_k$ is the estimated value of the $k$-th parameter and 
$\text{se}(\hat{\beta}_k)$ is the corresponding standard error. For example, if a researcher wants to ensure that all parameter estimates are significantly different from zero with 95\% confidence, then $|t|_{\text{min}}$ should be larger than 1.96.
In our study, we calculated the expected $|t|_{\text{min}}$ by averaging the minimum absolute $t$-ratios from $N=1000$ simulated datasets, which can be expressed as
\begin{equation}\label{eq:E t-ratio}
   E(|t|_{\text{min}}) =  \frac{1}{N} \sum_{n=1}^{N} |t|_{\text{min}}^{n},
\end{equation}
where $|t|_{\text{min}}^{n}$ is the minimum absolute $t$-ratio obtained from the $n$-th simulation dataset.
\begin{figure}[h]
    \centering
    \includegraphics[width=0.8\linewidth]{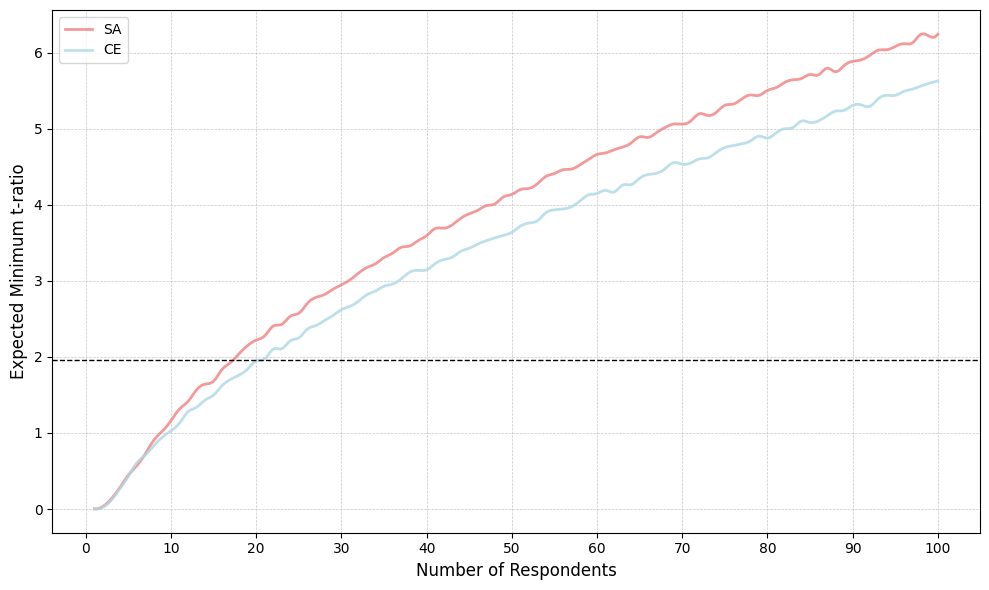}
    \caption{Expected minimum absolute $t$-ratio per sample size for the CE and SA designs.}
    \label{fig:t-ratio}
\end{figure}

Figures \ref{fig:t-ratio} illustrates the changes in the expected minimum absolute $t$-ratios of the CE and SA designs as the sample size increases. The dashed horizontal line represents the $t$-ratio of 1.96. When the number of respondents exceeds 7, the SA design yields higher expected minimum absolute $t$-ratios compared to the CE design, indicating that, with the same number of respondents, the SA design is more likely to achieve statistically significant parameter estimates. Specifically, the SA design requires only 18 respondents to state that all parameter estimates differ from zero with 95\% confidence, while the CE design requires 21 respondents.

\section{Discussion and conclusion}\label{Sec:conclusion}
In this study, we introduce for the first time the SA algorithm for generating Bayesian $\mathcal{D}$-optimal designs for DCEs. 
Our SA algorithm starts with a random choice design. In each iteration, it randomly generates a new choice design. The algorithm accepts not only all superior designs but also, with a specific probability, less optimal designs, thus effectively preventing early convergence. Compared to the well-known CE algorithm, our SA algorithm is more likely to explore a broader solution space and obtain better designs as output. 
Our computational experiments validate the superior performance of the SA algorithm across various scenarios, particularly when the prior preference information is very uncertain, or when there are relatively many prior parameters that are large in absolute size. In such instances, due to increased uncertainty and complexity in the optimization problem, the SA algorithm's advantage of avoiding premature convergence to local optima compared to the CE algorithm becomes more pronounced.

Our study highlights the potential of the SA algorithm for Bayesian optimal design in traditional DCEs. However, further research is required to adapt it for more complex experimental settings. In particular, two important extensions that require investigation are partial profile designs and mixture choice designs. 

Partial profile designs aim to mitigate the cognitive burden associated with choice tasks involving a large number of attributes. 
While the classical random utility model assumes that individuals tend to make compensatory decisions, empirical evidence suggests that respondents may adopt non-compensatory strategies due to the complexity of the decision-making task. An increase in the number of attributes increases the cognitive effort required, which may cause respondents to resort to simpler decision-making strategies, as indicated by \cite{caussade2005assessing}. To reduce the complexity of the comparisons and deter respondents from adopting simpler strategies, one approach is to maintain constant levels for some attributes in every choice situation. This approach leads to a partial profile design, as discussed by \cite{chrzan2010using}. 
When constructing such designs, the iterative process of the SA algorithm must ensure that a fixed number of attributes remain constant within each choice set. This  restriction necessitates the development of an efficient exploration rule to search the constrained design space.
Future modifications to the SA algorithm could enable it to incorporate the constraints required to construct partial profile designs.

Mixture choice designs involve attributes represented as proportions of ingredients in a mixture. For example, in the cocktail taste experiment conducted by \cite{courcoux1997methode}, researchers investigated how the proportions of three cocktail ingredients—mango juice, blackcurrant syrup, and lemon juice—affect individuals' preferences for the cocktail.
One major challenge in applying the SA algorithm to mixture choice designs lies in the inherent constraint that the proportions of all ingredients must sum to one. This constraint implies that whenever the proportion of one ingredient is altered during the iterative process, at least one other ingredient's proportion must be adjusted accordingly. To address this, the exploration rule in our current SA algorithm must be further refined to ensure an effective and comprehensive search within the constrained design space.
Notably, this modified SA algorithm could extend beyond mixture choice experiments to broader applications, including mixture experiments based on linear or generalized linear models \citep{cornell1988analyzing} and ecological studies using diversity–interaction modelling \citep{Kirwan2009Diversity}.

Finally, in future research, we recommend exploring alternative optimization methods for constructing Bayesian optimal designs in DCEs. Despite significant advances in the broader field of experimental design, the application of many well-developed algorithms to DCEs remains limited. Notable examples include the genetic algorithm \citep{WAGER2003293}, particle swarm optimization \citep{ruseckaite2017}, ant colony optimization \citep{BORROTTI2016259}, and variable neighborhood search \citep{GOOS2020201}. Conducting a detailed comparative analysis of these methods could offer valuable insights and practical guidance for practitioners in the field of DCEs.

\bibliography{ref}

\begin{thebibliography}{}

\bibitem[Aarts et~al., 1988]{aarts1988quantitative}
Aarts, E.~H., Korst, J.~H., and van Laarhoven, P.~J. (1988).
\newblock A quantitative analysis of the simulated annealing algorithm: A case study for the traveling salesman problem.
\newblock {\em Journal of Statistical Physics}, 50:187--206.

\bibitem[Angelis et~al., 2001]{angelis2001optimal}
Angelis, L., Bora-Senta, E., and Moyssiadis, C. (2001).
\newblock Optimal exact experimental designs with correlated errors through a simulated annealing algorithm.
\newblock {\em Computational Statistics \& Data Analysis}, 37(3):275--296.

\bibitem[Bliemer and Rose, 2010]{BLIEMER2010720}
Bliemer, M.~C. and Rose, J.~M. (2010).
\newblock Construction of experimental designs for mixed logit models allowing for correlation across choice observations.
\newblock {\em Transportation Research Part B: Methodological}, 44(6):720--734.
\newblock Methodological Advancements in Constructing Designs and Understanding Respondent Behaviour Related to Stated Preference Experiments.

\bibitem[Bliemer and Rose, 2011]{BLIEMER201163}
Bliemer, M.~C. and Rose, J.~M. (2011).
\newblock Experimental design influences on stated choice outputs: An empirical study in air travel choice.
\newblock {\em Transportation Research Part A: Policy and Practice}, 45(1):63--79.

\bibitem[Bohachevsky et~al., 1986]{bohachevsky1986generalized}
Bohachevsky, I.~O., Johnson, M.~E., and Stein, M.~L. (1986).
\newblock Generalized simulated annealing for function optimization.
\newblock {\em Technometrics}, 28(3):209--217.

\bibitem[Borrotti et~al., 2016]{BORROTTI2016259}
Borrotti, M., Minervini, G., {De Lucrezia}, D., and Poli, I. (2016).
\newblock Naïve bayes ant colony optimization for designing high dimensional experiments.
\newblock {\em Applied Soft Computing}, 49:259--268.

\bibitem[Bridges et~al., 2011]{BRIDGES2011403}
Bridges, J.~F., Hauber, A.~B., Marshall, D., Lloyd, A., Prosser, L.~A., Regier, D.~A., Johnson, F.~R., and Mauskopf, J. (2011).
\newblock Conjoint analysis applications in health—a checklist: A report of the ispor good research practices for conjoint analysis task force.
\newblock {\em Value in Health}, 14(4):403--413.

\bibitem[Burkard and Rendl, 1984]{BurkardRendl1984}
Burkard, R.~E. and Rendl, F. (1984).
\newblock A thermodynamically motivated simulation procedure for combinatorial optimization problems.
\newblock {\em European Journal of Operational Research}, 17(2):169--174.

\bibitem[Caussade et~al., 2005]{caussade2005assessing}
Caussade, S., Ort{\'u}zar, J. d.~D., Rizzi, L., and Hensher, D. (2005).
\newblock Assessing the influence of design dimensions on stated choice experiment estimates.
\newblock {\em Transportation Research Part B: Methodological}, 39(7):621--640.

\bibitem[Chrzan, 2010]{chrzan2010using}
Chrzan, K. (2010).
\newblock Using partial profile choice experiments to handle large numbers of attributes.
\newblock {\em International Journal of Market Research}, 52(6):827--840.

\bibitem[Connolly, 1990]{Connolly1990}
Connolly, D.~T. (1990).
\newblock An improved annealing scheme for the qap.
\newblock {\em European Journal of Operational Research}, 46(1):93--100.

\bibitem[Cook and Nachtsheim, 1980]{cook1980comparison}
Cook, R.~D. and Nachtsheim, C.~J. (1980).
\newblock A comparison of algorithms for constructing exact {D}-optimal designs.
\newblock {\em Technometrics}, 22(3):315--324.

\bibitem[Cornell, 1988]{cornell1988analyzing}
Cornell, J. (1988).
\newblock Analyzing data from mixture experiments containing process variables: A split-plot approach.
\newblock {\em Journal of Quality Technology}, 20(1):2--23.

\bibitem[Courcoux and S{\'e}m{\'e}nou, 1997]{courcoux1997methode}
Courcoux, P. and S{\'e}m{\'e}nou, M. (1997).
\newblock Une m\'ethode de segmentation pour l'analyse de donn\'ees issues de comparaisons par paires.
\newblock {\em Revue de statistique appliqu{\'e}e}, 45(2):59--69.

\bibitem[{de Bekker-Grob} et~al., 2019]{DEBEKKERGROB20191050}
{de Bekker-Grob}, E.~W., Swait, J.~D., Kassahun, H.~T., Bliemer, M.~C., Jonker, M.~F., Veldwijk, J., Cong, K., Rose, J.~M., and Donkers, B. (2019).
\newblock Are healthcare choices predictable? the impact of discrete choice experiment designs and models.
\newblock {\em Value in Health}, 22(9):1050--1062.

\bibitem[Dueck, 1993]{Dueck1993}
Dueck, G. (1993).
\newblock New optimization heuristics: the great deluge algorithm and the record-to-record travel.
\newblock {\em Journal of Computational Physics}, 104(1):86--92.

\bibitem[Franzin and St{\"u}tzle, 2019]{franzin2019revisiting}
Franzin, A. and St{\"u}tzle, T. (2019).
\newblock Revisiting simulated annealing: A component-based analysis.
\newblock {\em Computers \& Operations Research}, 104:191--206.

\bibitem[Geman and Geman, 1984]{geman1984stochastic}
Geman, S. and Geman, D. (1984).
\newblock Stochastic relaxation, {G}ibbs distributions, and the {B}ayesian restoration of images.
\newblock {\em IEEE Transactions on Pattern Analysis and Machine Intelligence}, 6(6):721--741.

\bibitem[Goos and Jones, 2011]{goos2011optimal}
Goos, P. and Jones, B. (2011).
\newblock {\em Optimal design of experiments: a case study approach}.
\newblock John Wiley \& Sons.

\bibitem[Goos et~al., 2020]{GOOS2020201}
Goos, P., Syafitri, U., Sartono, B., and Vazquez, A. (2020).
\newblock A nonlinear multidimensional knapsack problem in the optimal design of mixture experiments.
\newblock {\em European Journal of Operational Research}, 281(1):201--221.

\bibitem[Gotwalt et~al., 2009]{gotwalt2009fast}
Gotwalt, C.~M., Jones, B.~A., and Steinberg, D.~M. (2009).
\newblock Fast computation of designs robust to parameter uncertainty for nonlinear settings.
\newblock {\em Technometrics}, 51(1):88--95.

\bibitem[Huber and Zwerina, 1996]{huber1996importance}
Huber, J. and Zwerina, K. (1996).
\newblock The importance of utility balance in efficient choice designs.
\newblock {\em Journal of Marketing Research}, 33(3):307--317.

\bibitem[Hussin and St{\"u}tzle, 2014]{HussinStutzle2014}
Hussin, M.~S. and St{\"u}tzle, T. (2014).
\newblock Tabu search vs. simulated annealing for solving large quadratic assignment instances.
\newblock {\em Computers \& Operations Research}, 43:286--291.

\bibitem[Johnson et~al., 1989]{Johnson1989}
Johnson, D.~S., Aragon, C.~R., McGeoch, L.~A., and Schevon, C. (1989).
\newblock Optimization by simulated annealing: An experimental evaluation: Part i, graph partitioning.
\newblock {\em Operations Research}, 37(6):865--892.

\bibitem[Kessels et~al., 2006]{kessels2006comparison}
Kessels, R., Goos, P., and Vandebroek, M. (2006).
\newblock A comparison of criteria to design efficient choice experiments.
\newblock {\em Journal of Marketing Research}, 43(3):409--419.

\bibitem[Kessels et~al., 2008]{kessels2008recommendations}
Kessels, R., Jones, B., Goos, P., and Vandebroek, M. (2008).
\newblock Recommendations on the use of {B}ayesian optimal designs for choice experiments.
\newblock {\em Quality and Reliability Engineering International}, 24(6):737--744.

\bibitem[Kessels et~al., 2009]{kessels2009efficient}
Kessels, R., Jones, B., Goos, P., and Vandebroek, M. (2009).
\newblock An efficient algorithm for constructing {B}ayesian optimal choice designs.
\newblock {\em Journal of Business and Economic Statistics}, 27(2):279--291.

\bibitem[Kessels et~al., 2011]{Kessels2011usefulness}
Kessels, R., Jones, B., Goos, P., and Vandebroek, M. (2011).
\newblock The usefulness of {B}ayesian optimal designs for discrete choice experiments.
\newblock {\em Applied Stochastic Models in Business and Industry}, 27(3):173--188.

\bibitem[Kirkpatrick et~al., 1983]{kirkpatrick1983optimization}
Kirkpatrick, S., Gelatt~Jr, C.~D., and Vecchi, M.~P. (1983).
\newblock Optimization by simulated annealing.
\newblock {\em Science}, 220(4598):671--680.

\bibitem[Kirwan et~al., 2009]{Kirwan2009Diversity}
Kirwan, L., Connolly, J., Finn, J.~A., Brophy, C., Lüscher, A., Nyfeler, D., and Sebastià, M.-T. (2009).
\newblock Diversity–interaction modeling: estimating contributions of species identities and interactions to ecosystem function.
\newblock {\em Ecology}, 90(8):2032--2038.

\bibitem[Liu et~al., 2006]{liu2006improved}
Liu, A., Wang, J., Han, G., Wang, S., and Wen, J. (2006).
\newblock Improved simulated annealing algorithm solving for 0/1 knapsack problem.
\newblock In {\em Sixth International Conference on Intelligent Systems Design and Applications}, volume~2, pages 1159--1164. IEEE.

\bibitem[Liu et~al., 2024]{Liu2024}
Liu, J., Kassas, B., and Lai, J. (2024).
\newblock Investigating the role of political messaging on preferences for local food products in the united states.
\newblock {\em Journal of Agricultural and Applied Economics}, page 1–24.

\bibitem[Lundy and Mees, 1986]{LundyMees1986}
Lundy, M. and Mees, A. (1986).
\newblock Convergence of an annealing algorithm.
\newblock {\em Mathematical Programming}, 34(1):111--124.

\bibitem[Luyten et~al., 2015]{Luyten2015}
Luyten, J., Kessels, R., Goos, P., and Beutels, P. (2015).
\newblock Public preferences for prioritizing preventive and curative health care interventions: A discrete choice experiment.
\newblock {\em Value in Health}, 18(2):224--233.

\bibitem[Malek et~al., 1989]{malek1989serial}
Malek, M., Guruswamy, M., Pandya, M., and Owens, H. (1989).
\newblock Serial and parallel simulated annealing and tabu search algorithms for the traveling salesman problem.
\newblock {\em Annals of Operations Research}, 21:59--84.

\bibitem[Metropolis et~al., 1953]{metropolis1953equation}
Metropolis, N., Rosenbluth, A.~W., Rosenbluth, M.~N., Teller, A.~H., and Teller, E. (1953).
\newblock Equation of state calculations by fast computing machines.
\newblock {\em The Journal of Chemical Physics}, 21(6):1087--1092.

\bibitem[Meyer and Nachtsheim, 1988]{meyer1988constructing}
Meyer, R.~K. and Nachtsheim, C.~J. (1988).
\newblock Constructing exact {D}-optimal experimental designs by simulated annealing.
\newblock {\em American Journal of Mathematical and Management Sciences}, 8(3-4):329--359.

\bibitem[Meyer and Nachtsheim, 1995]{meyer1995coordinate}
Meyer, R.~K. and Nachtsheim, C.~J. (1995).
\newblock The coordinate-exchange algorithm for constructing exact optimal experimental designs.
\newblock {\em Technometrics}, 37:60--69.

\bibitem[Monahan and Genz, 1997]{monahan1997spherical}
Monahan, J. and Genz, A. (1997).
\newblock Spherical-radial integration rules for {B}ayesian computation.
\newblock {\em Journal of the American Statistical Association}, 92(438):664--674.

\bibitem[Mysovskikh, 1980]{mysovskikh1980approximation}
Mysovskikh, I.~P. (1980).
\newblock The approximation of multiple integrals by using interpolatory cubature formulae.
\newblock In {\em Quantitative Approximation}, pages 217--243. Academic Press.

\bibitem[Qian and Ding, 2007]{qian2007simulated}
Qian, F. and Ding, R. (2007).
\newblock Simulated annealing for the 0/1 multidimensional knapsack problem.
\newblock {\em Numerical Mathematics: A Journal of Chinese Universities (English Series)}, 16(4):320.

\bibitem[Rose and Bliemer, 2013]{rose2013sample}
Rose, J.~M. and Bliemer, M.~C. (2013).
\newblock Sample size requirements for stated choice experiments.
\newblock {\em Transportation}, 40(6):1021--1041.

\bibitem[Rossi and Allenby, 2003]{Rossi2003}
Rossi, P.~E. and Allenby, G.~M. (2003).
\newblock {B}ayesian statistics and marketing.
\newblock {\em Marketing Science}, 22(3):304--328.

\bibitem[Ruseckaite et~al., 2017]{ruseckaite2017}
Ruseckaite, A., Goos, P., and Fok, D. (2017).
\newblock {B}ayesian {D}-optimal choice designs for mixtures.
\newblock {\em Journal of the Royal Statistical Society. Series C (Applied Statistics)}, 66(2):363--386.

\bibitem[S\'andor and Wedel, 2001]{sandor2001designing}
S\'andor, Z. and Wedel, M. (2001).
\newblock Designing conjoint choice experiments using managers' prior beliefs.
\newblock {\em Journal of Marketing Research}, 38(4):430--444.

\bibitem[Strenski and Kirkpatrick, 1991]{strenski1991analysis}
Strenski, P.~N. and Kirkpatrick, S. (1991).
\newblock Analysis of finite length annealing schedules.
\newblock {\em Algorithmica}, 6(1-6):346--366.

\bibitem[Szu and Hartley, 1987]{SzuHartley1987}
Szu, H. and Hartley, R. (1987).
\newblock Fast simulated annealing.
\newblock {\em Physics Letters A}, 122(3):157--162.

\bibitem[Tam, 1992]{Tam1992}
Tam, K.~Y. (1992).
\newblock A simulated annealing algorithm for allocating space to manufacturing cells.
\newblock {\em International Journal of Production Research}, 30(1):63--87.

\bibitem[Tian and Yang, 2017]{Tian2017Efficiency}
Tian, T. and Yang, M. (2017).
\newblock Efficiency of the coordinate-exchange algorithm in constructing exact optimal discrete choice experiments.
\newblock {\em Journal of Statistical Theory and Practice}, 11(2):254--268.

\bibitem[Train, 2009]{Train2009}
Train, K.~E. (2009).
\newblock {\em Discrete Choice Methods with Simulation}.
\newblock Cambridge University Press, Cambridge.

\bibitem[{Van Acker} et~al., 2020]{VANACKER2020759}
{Van Acker}, V., Kessels, R., {Palhazi Cuervo}, D., Lannoo, S., and Witlox, F. (2020).
\newblock Preferences for long-distance coach transport: Evidence from a discrete choice experiment.
\newblock {\em Transportation Research Part A: Policy and Practice}, 132:759--779.

\bibitem[Wager and Nichols, 2003]{WAGER2003293}
Wager, T.~D. and Nichols, T.~E. (2003).
\newblock Optimization of experimental design in fmri: a general framework using a genetic algorithm.
\newblock {\em NeuroImage}, 18(2):293--309.

\bibitem[Yu et~al., 2008]{yu2008model}
Yu, J., Goos, P., and Vandebroek, M. (2008).
\newblock Model-robust design of conjoint choice experiments.
\newblock {\em Communications in Statistics—Simulation and Computation}, 37(8):1603--1621.

\bibitem[Yu et~al., 2010]{yu2010comparing}
Yu, J., Goos, P., and Vandebroek, M. (2010).
\newblock Comparing different sampling schemes for approximating the integrals involved in the efficient design of stated choice experiments.
\newblock {\em Transportation Research Part B: Methodological}, 44(10):1268--1289.

\end{thebibliography}
\clearpage

\appendix
\renewcommand{\thesection}{Appendix \Alph{section}.}
\section{Tables}\label{sec:app1}
\renewcommand{\thetable}{A\arabic{table}}
\setcounter{table}{0}
\begin{table}[h!]
    \centering
     \caption{Average $\mathcal{D}_B$ values, runtimes, and relative $\mathcal{D}_B$-efficiencies for the CE and SA algorithms across different $\lambda$ and $\kappa$ values based on 100 random initial designs with 3 profiles per choice set.}
    \begin{tabular}{llcccccc}
     \hline
\multirow{2}{*}{$\lambda$} & \multirow{2}{*}{$\kappa$}& \multicolumn{2}{c}{CE}&  &\multicolumn{2}{c}{SA}&\multirow{2}{*}{Avg. $\mathcal{D}_{B}\text{-eff}({\bf {X}}_{CE},{\bf {X}}_{SA})$}\\
\cline{3-4} \cline{6-7}
  & & Avg. $\mathcal{D}_B$& Avg. runtime (s)& & Avg. $\mathcal{D}_B$& Avg. runtime (s)& \\
  \hline
           1&  1
&  \phantom{1}7.82 &  14.91 
&&  \phantom{1}8.17 &  \phantom{1}82.03 
& 96.15\%\\
           1&  0.5&  15.70 &  15.64 
&&  16.06 &  121.43 
& 96.13\%\\
           1&  0.33&  16.72 &  15.33 
&&  17.24 &  157.50 
& 94.47\%\\
           0.5&  1
&  \phantom{1}9.02 &  15.19 
&&  \phantom{1}9.23 &  \phantom{1}87.71 
& 97.63\%\\
           0.5&  0.5&  17.38 &  17.90 
&&  17.47 &  129.59 
& 98.96\%\\
           0.5&  0.33&  18.59 &  20.76 
&&  18.73 &  136.69 
& 98.50\%\\
           0.33&  1
&  \phantom{1}9.38 &  19.62 
&&  \phantom{1}9.58 &  \phantom{1}70.47 
& 97.82\%\\
           0.33&  0.5&  18.03 &  12.64 
&&  18.11 &  115.14 
& 99.14\%\\
           0.33&  0.33&  19.31 &  12.93 
&&  19.39 &  118.32 
& 99.03\%\\
           \hline
    \end{tabular}
   
    \label{tab:CE_VS_SA_3}
\end{table}
\begin{longtable}{c c c c c c c c  c c c c c c}
\caption{$\mathcal{D}_B$-optimal designs generated by the CE and SA algorithms containing 15 choice sets of 2 profiles.}
\label{tab:design2}\\
\toprule
Choice set&  \multicolumn{6}{c}{CE} & & \multicolumn{6}{c}{SA}\\
\midrule
1&  1&  2&  2&  3&  4&  4
&  &  1&  2& 2& 2& 3&2
\\
         1&  2&  2&  2&  1&  2&  4
&  &  1&  2& 2& 3& 1&1
\\
         2&  2&  1&  2&  2&  1&  1
&  &  2&  2& 1& 1& 3&1
\\
         2&  2&  2&  1&  3&  2&  1
&  &  2&  2& 1& 3& 1&3
\\
         3&  1&  2&  1&  4&  3&  2
&  &  2&  2& 2& 2& 2&4
\\
         3&  1&  2&  1&  3&  2&  3
&  &  2&  2& 2& 3& 3&2
\\
         4&  1&  1&  1&  1&  3&  2
&  &  1&  2& 2& 1& 2&3
\\
         4&  1&  1&  1&  2&  3&  1
&  &  2&  2& 1& 3& 1&2
\\
         5&  2&  1&  1&  1&  3&  4
&  &  2&  1& 1& 1& 3&3
\\
 5& 2& 2& 1& 2& 1& 2
& & 2& 2& 1& 2& 2&1
\\
 6& 1& 2& 2& 1& 4& 1
& & 1& 2& 1& 2& 4&3
\\
 6& 1& 2& 1& 2& 3& 3
& & 2& 2& 2& 1& 4&1
\\
 7& 2& 1& 2& 4& 2& 1
& & 1& 2& 1& 1& 1&3
\\
 7& 1& 1& 2& 4& 1& 3
& & 1& 1& 1& 3& 2&2
\\
 8& 2& 1& 1& 3& 4& 3
& & 1& 2& 2& 2& 3&3
\\
 8& 1& 2& 2& 1& 3& 4
& & 1& 2& 2& 3& 4&4
\\
 9& 2& 1& 2& 3& 3& 2
& & 2& 2& 1& 1& 4&1
\\
 9& 2& 2& 2& 4& 1& 3
& & 1& 2& 2& 3& 3&1
\\
 10& 2& 1& 1& 4& 1& 3
& & 2& 2& 2& 3& 2&2
\\
 10& 2& 2& 1& 1& 1& 1
& & 1& 2& 2& 4& 4&2
\\
 11& 1& 2& 1& 4& 2& 1
& & 2& 1& 1& 4& 2&3
\\
 11& 2& 2& 1& 2& 3& 2
& & 1& 1& 1& 3& 3&4
\\
 12& 2& 2& 1& 2& 1& 4
& & 1& 1& 2& 2& 3&1
\\
 12& 2& 2& 1& 3& 3& 1
& & 1& 1& 2& 1& 2&2
\\
 13& 1& 1& 2& 3& 2& 2
& & 1& 1& 1& 3& 2&3
\\
 13& 1& 1& 2& 1& 3& 3
& & 1& 1& 2& 2& 1&2
\\
 14& 2& 2& 2& 3& 1& 4
& & 2& 1& 2& 2& 4&3
\\
 14& 2& 2& 1& 2& 4& 4
& & 1& 2& 1& 4& 3&4
\\
 15& 1& 1& 2& 2& 2& 3
& & 2& 1& 1& 4& 3&4
\\
 15& 2& 1& 1& 3& 1& 1& & 1& 2& 2& 3& 4&2\\
 \bottomrule
\end{longtable}
\clearpage
\section{Figures}
\renewcommand{\thefigure}{B\arabic{figure}}
\setcounter{figure}{0}
\begin{figure}[H] 
    \centering
    \begin{subfigure}{0.3\textwidth}
        \includegraphics[width=\linewidth]{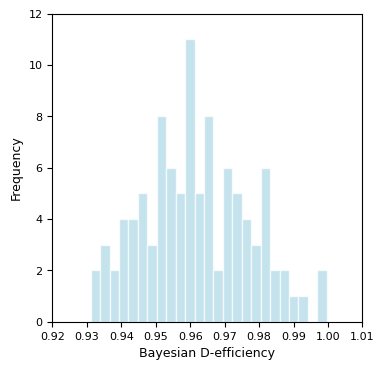} 
        \caption{$\lambda = 1, \kappa = 1$}
    \end{subfigure}
    \begin{subfigure}{0.3\textwidth}
        \includegraphics[width=\linewidth]{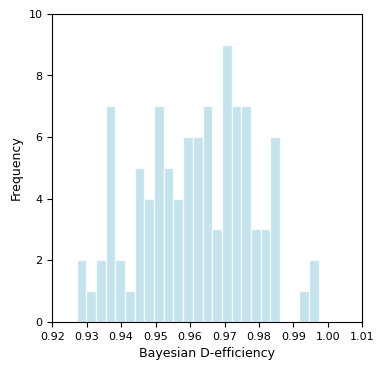}
        \caption{$\lambda = 1, \kappa = 1/2$}
    \end{subfigure}
    \begin{subfigure}{0.3\textwidth}
        \includegraphics[width=\linewidth]{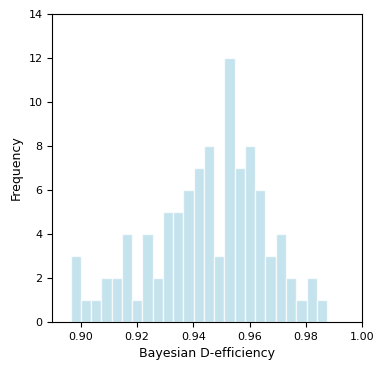}
        \caption{$\lambda = 1, \kappa = 1/3$}
    \end{subfigure}

    \begin{subfigure}{0.3\textwidth}
        \includegraphics[width=\linewidth]{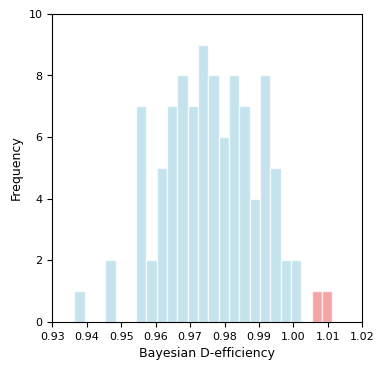}
        \caption{$\lambda = 1/2, \kappa = 1$}
    \end{subfigure}
    \begin{subfigure}{0.3\textwidth}
        \includegraphics[width=\linewidth]{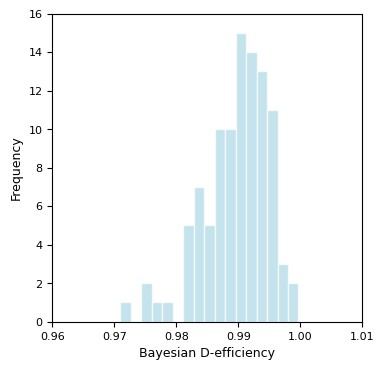}
        \caption{$\lambda =1/2, \kappa =1/2$}
    \end{subfigure}
    \begin{subfigure}{0.3\textwidth}
        \includegraphics[width=\linewidth]{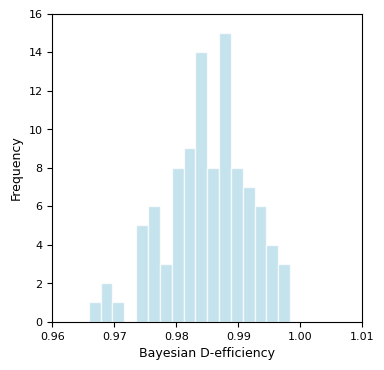}
        \caption{$\lambda = 1/2, \kappa = 1/3$}
    \end{subfigure}

    \begin{subfigure}{0.3\textwidth}
        \includegraphics[width=\linewidth]{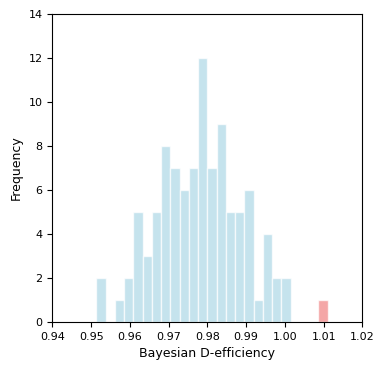}
        \caption{$\lambda = 1/3, \kappa = 1$}
    \end{subfigure}
    \begin{subfigure}{0.3\textwidth}
        \includegraphics[width=\linewidth]{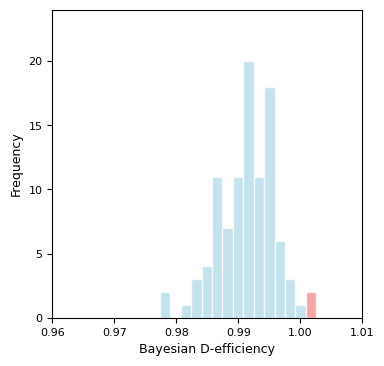}
        \caption{$\lambda = 1/3, \kappa = 1/2$}
    \end{subfigure}
    \begin{subfigure}{0.3\textwidth}
        \includegraphics[width=\linewidth]{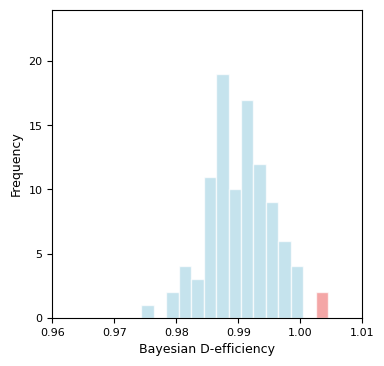}
        \caption{$\lambda = 1/3, \kappa = 1/3$}
    \end{subfigure}
    
   \caption{Histograms of Bayesian $\mathcal{D}_{B}$-efficiencies of CE optimal designs relative to SA optimal designs (or $\mathcal{D}_{B}\text{-eff}({\bf {X}}_{CE},{\bf {X}}_{SA})$) across nine different prior distributions, where the designs consist of 3 profiles per choice set.}
\label{fig:CE_VS_SA_3}
\end{figure}

\end{document}